\documentclass[aps,pra,twocolumn,showpacs,longbibliography]{revtex4}
\usepackage{graphicx}
\usepackage{hyperref}
\usepackage{bm,color}
\usepackage{dcolumn}
\usepackage{epsfig,amsbsy,bm,amssymb}
\usepackage{tabu}
\usepackage{tabularx}
\renewcommand*{\[}{\begin{equation}}
\renewcommand*{\]}{\end{equation}}

\renewcommand{\k}{{k}}
\newcommand{\hs}{\hspace*}
\newcommand{\vs}{\vspace*}
\newcommand{\np}{\newpage}

\newcommand{\eref}[1] {(\ref{#1})}
\newcommand{\Eref}[1] {Eq.~(\ref{#1})}
\newcommand{\Fref}[1] {Fig. \ref{#1}}
\newcommand{\Tref}[1] {Table \ref{#1}}
\newcommand{\ra}{\rangle}
\newcommand{\la}{\langle}
\newcommand{\nn}{\nonumber}
\newcommand{\be}{\begin{equation}}
\newcommand{\ee}{\end{equation}}
\newcommand{\br}{\begin{eqnarray*}}
\newcommand{\er}{\end{eqnarray*}}
\newcommand{\ba}{\begin{eqnarray}}
\newcommand{\ea}{\end{eqnarray}}
\newcommand{\bp}{\begin{minipage}}
\newcommand{\ep}{\end{minipage}}
\newcommand{\bt}{\begin{tabular}}
\newcommand{\et}{\end{tabular}}
\newcommand{\ds}{\displaystyle}
\newcommand{\bs}{\bigskip}
\newcommand{\e}{\epsilon}

\renewcommand{\r}{{\bm r}}
\renewcommand{\k}{{\bm k}}

\renewcommand{\b}{\bf}
\renewcommand{\b}{\beta}
\newcommand{\w}{\omega}

\begin{document}
\title{Simulation of angular resolved RABBITT measurements in noble
  gas atoms}

\date{\today}

\author{Alexander W. Bray, Faiza Naseem and  Anatoli~S.~Kheifets}
\bs

\affiliation{Research School of Physics and Engineering, The
Australian National University, Canberra ACT 0200, Australia}



\begin{abstract}
We simulate angular resolved RABBITT (Reconstruction of Attosecond
Beating By Interference of Two-photon Transitions) measurements on 
valence shells of noble gas atoms (Ne, Ar, Kr, and Xe). Our non-perturbative
numerical simulation is based on solution of the time-dependent
Schr\"odinger equation for a target atom driven by an ionizing XUV and
dressing IR fields.  From these simulations we extract the angular
dependent magnitude and phase of the RABBITT oscillations and deduce
the corresponding angular anisotropy $\beta$ parameter and Wigner time
delay $\tau_W$ for the single XUV photon absorption which initiates
the RABBITT process. Said $\beta$ and $\tau_W$ parameters are
compared with calculations in the random phase approximation with exchange (RPAE)
which includes inter-shell correlation. This comparison is used to
test various  effective  potentials employed in the one-electron TDSE.
In lighter atoms (Ne and Ar), several effective potentials are found
to provide accurate simulation of RABBITT measurements for a wide range of photon
energies up to 100~eV above the valence shell threshold. In heavier
atoms (Kr and Xe), the onset of strong correlation with the $d$-shell 
restricts the validity of the single active electron approximation to
several tens of eV above the valence shell threshold.
\end{abstract}

\maketitle

\section{Introduction}

Angular resolved RABBITT (Reconstruction of Attosecond Beating By
Interference of Two-photon Transitions) experiments have been used to
coherently control the photoelectron emission direction
\cite{PhysRevLett.109.083001} and, more recently, to measure angular
dependent time delay in atomic photoionization
\cite{PhysRevA.94.063409,Cirelli2018}.  These experiments bring
sensitive information on ultrafast electron dynamics influenced by
correlation and exchange effects. Theoretical modeling of the angular
resolved RABBITT process have been provided within the framework of
lowest order perturbation theory (LOPT)
\cite{0953-4075-47-12-124012,0953-4075-50-15-154002} and
non-perturbatively, by solving the time-dependent Schr\"odinger
equation (TDSE) \cite{PhysRevA.94.063409,PhysRevA.96.013408}.
As in our preceding paper \cite{PhysRevA.96.013408}, we solved TDSE
for a noble gas atom (He and Ne) driven by an ionizing XUV and
dressing IR fields in the configuration of a typical RABBITT
measurement. From this solution we deduced the angular dependence of
the photoemission time delay as measured by the RABBITT technique
\cite{MullerAPB2002,TomaJPB2002}. Our model was calibrated against a
recent angular resolved measurement on He \cite{PhysRevA.94.063409}.
We employed the soft photon approximation (SPA) and used a hydrogenic
continuum-continuum (CC) correction to connect the magnitude and phase
of the RABBITT oscillations with the angular anisotropy $\beta$
parameter and the Wigner time delay $\tau_W$ for the single XUV photon
absorption which initiates the RABBITT process.

Solution of the TDSE in \cite{PhysRevA.96.013408} was obtained in the
single active electron (SAE) approximation and utilized the optimized
effective potentials (OEP) of \citet{Sarsa2004163}. While such
approach was found to be valid for He, this remains to be shown for Ne
and heavier noble gas atoms. In the present work, we conduct these
tests for noble gases from Ne to Xe by making comparison of the
$\beta$ and $\tau_W$ parameters with those coming from calculations
performed in the random phase approximation with exchange (RPAE), the
latter including inter-shell correlation and exchange of the
photoelectron with the remaining ionic core. These effects are not
included in the TDSE/SAE model. 
However, the latter model takes an accurate account of ultrafast
electron dynamics whereas the RPAE is unable to do so by its basis
based construction.
In lighter atoms (Ne and Ar), several effective potentials are found
to provide accurate simulation of RABBITT measurements over a wide
range of photon energies up to 100~eV above the valence shell
threshold. In heavier atoms (Kr and Xe), the onset of strong
correlation with the sub-valent $d$-shell restricts validity of the
SAE approximation to several tens of eV above the valence shell
threshold.

A further goal of the present work is to test universality of the
hydrogenic CC correction ($\tau_{cc}$). This correction relates the single-photon
Wigner time delay ($\tau_W$) and the measured atomic time delay ($\tau_a$) via
\be
\tau_a= \tau_W+\tau_{cc}  \ .
\label{atomic}
\ee
A hydrogenic CC correction was used in the theoretical analysis of the
photoemission time delay measured close to the $3s$ ionization cross-section 
minimum in Ar \cite{PhysRevA.85.053424}. The theoretical and
experimental time delays reported in \cite{PhysRevA.85.053424} differed
by as much as 50~as and no plausible explanation to this disagreement
was found to date.  We address this issue in the present work. 
More recently, the RABBITT measurement on Ne of
\citet{Isingerer7043} has finally reconciled the persistent
disagreement between the earlier experiment \cite{M.Schultze06252010}
and a large number of theoretical predictions
\cite{PhysRevLett.105.233002, PhysRevA.84.061404,PhysRevA.86.061402,
  PhysRevA.87.063404,PhysRevA.89.033417,PhysRevA.97.013422}. Our present
calculations are similarly in perfect agreement with \cite{Isingerer7043}.

The remainder of this paper is organized as follows. In Section \ref{theory}
we describe our method and numerical techniques.  In Section
\ref{results} we present and analyse our numerical data. We conclude by
highlighting links with existing experimental measurements and 
propose several new areas of interest.

\section{Theory}
\label{theory}

\subsection{Solution of TDSE}
\label{TDSE}

As previously \cite{PhysRevA.96.013408}, we solve the one-electron TDSE
for a target atom 
\begin{equation}
\label{TDSE}
i {\partial \Psi(\r) / \partial t}=
\left[\hat H_{\rm atom} + \hat H_{\rm int}(t)\right]
\Psi(\r) \ ,
\end{equation}
where the radial part of the atomic Hamiltonian
\be
\label{Hat}
\hat H_{\rm atom}(r) = 
-\frac12{d^2\over dr^2} +{l(l+1)\over 2r^2} + V(r)
\ee
contains an effective one-electron potential $V(r)$. The various potentials considered are
detailed in Sec.~\ref{Potential}.  The Hamiltonian $\hat H_{\rm
  int}(t)$ describes  interaction with the external field and is
written in the velocity gauge
\be
\label{gauge}
\hat H_{\rm int}(t) =
 {\bm A}(t)\cdot \hat{\bm p} \ \ , \ \ 
{\bm A(t)}=-\int_{0}^{t}{\bm F(t')}\ d t' \ .
\ee
This external field is comprised of both XUV and IR fields.
The XUV field is modelled by an attosecond pulse train (APT) with
the vector potential
\ba
\label{vectorGauss}
A_x(t) &=& \sum_{n=-5}^5 (-1)^n A_n \exp\left(
-2\ln2
{(t-nT/2)^2\over \tau_x^2}
\right)\nonumber\\
&&\times
\cos\Big[\omega_x(t-nT/2)\Big]  \ ,
\ea
where
$$
A_n = A_0
\exp\left(-2\ln2
{(nT/2)^2\over \tau_T^2}\right) \ .
$$
Here $A_0$ is the vector potential peak value
%
%
and $T=2\pi/\omega$ is the period of the IR field.
The XUV central frequency is $\omega_x$ and the 
time constants $\tau_x, \tau_T$ are chosen  to span a sufficient
number of harmonics in the range of photon frequencies of 
interest for a given atom.

The vector potential of the IR pulse is modelled by the cosine
squared envelope
\be
\label{vectorSin2}
A(t) = A_0 \cos^2
\left(
{\pi (t-\tau)\over 2\tau_{\rm IR}}
\right)
\cos[\omega(t-\tau)] \ .
\ee
The IR pulse is shifted relative to the APT by a variable delay $\tau$
such that the RABBITT signal of the even $2q$ sideband (SB) oscillates
as
\be
S_{2q}(\tau) =
A+B\cos[2\omega\tau-C]
\ .
\label{oscillation}
\ee
Solution of the TDSE \eref{TDSE} is found using the iSURF method
as given in \citet{0953-4075-49-24-245001}.
A typical calculation with XUV and IR field intensities of $5\times10^{9}$ and $3\times10^{10}$ W/cm$^2$ respectively
would take up to 35 CPU hours for each $\tau$.

The RABBITT parameters $A$, $B$ and $C$ entering
Eq.~(\ref{oscillation}) can be expressed via the absorption and
emission amplitudes
\ba
A&=&|{\cal M}^{(-)}_{\k}|^2+|{\cal M}^{*(+)}_{\k}|^2
\  , \ 
B=2{\rm Re} \left[{\cal M}^{(-)}_{\k}{\cal
    M}^{*(+)}_{\k}\right]
\nn\\
C&=& \arg\left[{M}^{(\rm -)}_{\k}{ M}^{*(\rm +)}_{\k}\right]
=2\w\tau_a \ .
\label{abc}
\ea
Here ${\cal M}^{(\pm)}_{\k}$ are complex amplitudes for the angle-resolved
photoelectron produced by adding or subtracting an IR photon,
respectively.  By adopting the soft photon approximation (SPA)
\cite{Maquet2007} we can write
\ba
\label{soft}
A,B &\propto&
|J_1({\bm \alpha}_0\cdot\k)|^2
|\la f|z|i \ra|^2\\
&\propto&
\left[1+\beta P_2(\cos\theta_k)\right]
\cos^2\theta_k  \ .
\nn
\ea
Here we made a linear approximation to the Bessel function as the
parameter $ \alpha_0= F_0/\omega^2$ is small in a weak IR field.
See Appendix for a more detailed derivation. In \Eref{soft} $\theta_k$
is the angle between the photoelectron emission direction $\hat\k$ and
the electric field vector of the linearly polarised light.  By fitting
the calculated angular dependence of the $A$ and $B$ parameters with
the SFA expression \Eref{soft} we can obtain the two sets of the
angular anisotropy parameters $\beta^{\rm SB}_A$ and $\beta^{\rm
  SB}_B$ and compare them with the value calculated by the RPAE
model. At the same time, we derive the angular dependence from the odd
high harmonic (HH) peaks by fitting angular variation of their
amplitude with $1+\beta^{\rm HH} P_2(\cos\theta_k)$. Thus, for each
target atom three sets of $\beta$ parameters are extracted and
analyzed over a wide photon energy range.

\citet{PhysRevLett.109.083001} proposed a different parameterization
of the angular dependence of the RABBITT signal. In case the APT has
only the odd HH peaks, it reads
\ba
\label{Laurent}
F_q(\theta_k,\tau) &=&
\sum\limits_{j=0}^{2L_{\max}}
\beta_j(q,\tau)P_j(\cos\theta_k) \\
&\propto&
1+\b_2P_2(\cos\theta_k)
+\b_4P_4(\cos\theta_k)
\nn
\ea
While $\beta_2$ in \Eref{Laurent} is identical with our definition of
$\beta^{\rm HH}$, $\beta_2$ and $\beta_4$ can be expressed via
$\beta^{\rm SB}$. By expanding \Eref{soft} over the Legendre
polynomials, we arrive to the following expressions:
\be
\label{b24}
\b_2={70+55\b^{\rm SB}\over 35+14\b^{\rm SB}}
\  \ \ ,  \ \ \
\b_4 = {36\b^{\rm SB}\over 35+14\b^{\rm SB}} \ .
\ee
In the following, we will show that in all presently studied cases,
$
\b^{\rm HH}\simeq\b^{\rm SB}_A\simeq\b^{\rm SB}_B
$
and one set of $\beta$ parameters fits all the RABBITT
measurement. The $\b_4$ and $\b_2$ parameters depend on this $\beta$
linearly. Thus $\b_4$ parameter is redundant and its introduction by
\citet{Cirelli2018} is superfluous.

The $C$ parameter is converted to the atomic time delay $\tau_a$ by
\eref{abc} and analyzed as a function of the photoelectron direction
relative to the polarization axis. The angular dependence of $\tau_a$
is compared with the analogous dependence of the Wigner time delay
$\tau_{\rm W}$ \cite{PhysRevA.94.013423}.  The time delay difference
$\tau_a-\tau_{\rm W}$ in the zero angle direction is compared with
the hydrogenic CC correction $\tau_{\rm CC}$ \cite{Dahlstrom2012}.

\subsection{One-electron potential}
\label{Potential}

\begin{figure}[h]
\epsfxsize=6.5cm
\epsffile{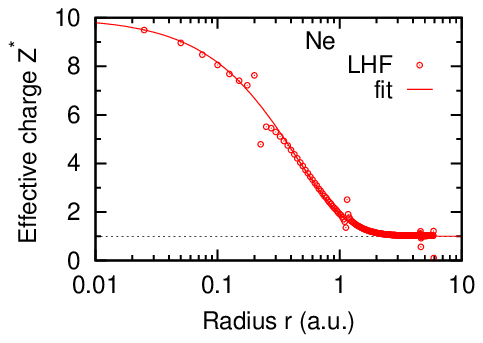}
\vs{9cm}
\caption{(Color online) The effective charge $Z^*(r) = -rV(r)$ of Ne
  generated from various one-electron potentials.  Top: the effective
  charge $Z^*$ derived from the LHF potential using the HF radial
  orbital $\kappa=0.01$ and $\ell=0$ is shown with the (red) open
  circles. The fit with the analytic expression \eref{ZHF} is shown
  with the (red) solid line.
Middle: the effective charges generated from the optimized effective
potential (OEP) of \cite{Sarsa2004163} and  the LHF potential are
compared with the spherically symmetric Hartree
potential \eref{Hartree}.
Bottom: The charge difference (exchange hole) $Z^*-Z^*_{\rm H}$ for the OEP
and LHF potentials.
\label{Fig1}} 
\end{figure}

\begin{figure}[th]

\epsfxsize=6.5cm
\epsffile{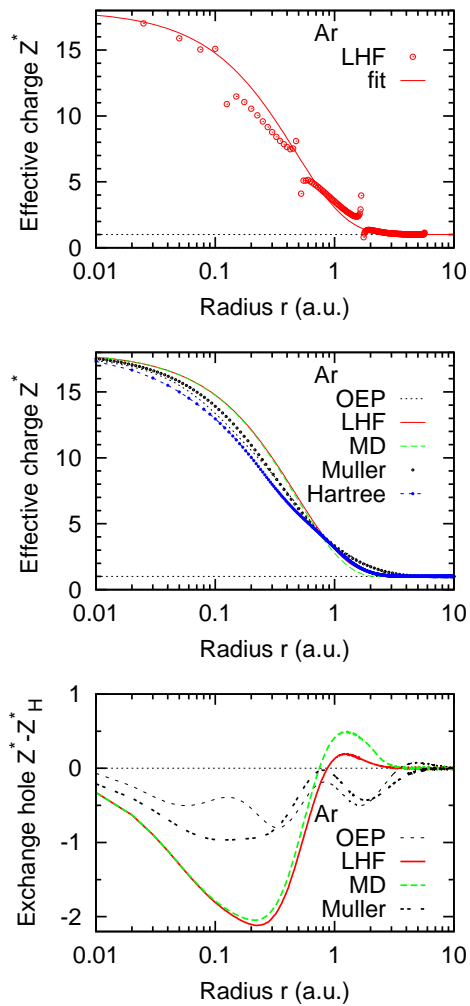}
\vs{9cm}

\caption{(Color online) Same as \Fref{Fig1} for argon. In addition,
  the effective charge $Z^*(r) = -rV(r)$ is generated from the atomic
  pseudopotential of \citet{MILLER197716} (labeled MD) and the Muller
  potential \cite{Muller2002}.
\label{Fig2}} 
\end{figure}

In our previous work on He and Ne, we employed an optimized effective
potential (OEP) \cite{Sarsa2004163}.  This potential is derived by a
simplified treatment of the exchange term in the Hartree-Fock (HF)
equations using the Slater X-$\alpha$  ansatz
\cite{PhysRev.81.385}. The OEP potential takes the form
\ba
\label{OEP}
V_e(r)&=& -{1\over r} 
\left(
1+(Z_0-1)
\sum_{p=0}^S
\sum_{k=1}^{n_p}
c_{k,p}r^pe^{-\beta_{k,p}r}
\right)
\nn\\&&
\equiv -{Z^*(r)\over r}
\ea
where the effective charge $Z^*(r)$ varies from the unscreened nucleus
charge $Z_0$ as $r\to0$ and unity at large distances $r\to\infty$. The
former limit is satisfied by imposing the condition
$\sum_{k=1}^{n_0} c_{k,0}=1$. The effective charges $Z^*(r)$ for Ne
and Ar are shown in \Fref{Fig1} and \Fref{Fig2}, respectively.


It is instructive to compare $Z^*$ with the effective charge derived
from the spherically symmetric part of the Hartree potential
$
Z^*_{\rm H} = Z_0-rV_{\rm H}(r)
$
where
$$
V_{\rm H}(r)={1\over 4\pi}
\int d\Omega_r
\int\limits_0^\infty d\r' \, {\rho(\r')
  \over |\r'-\r|}
\ \ ,  \
\rho(\r') = \sum_{nlm} |\psi_{nlm}(\r')|^2 \ .
$$
By way of spherical integration, the above expression can be reduced
to the following radial integral
\be
\label{Hartree}
V_{\rm H}(r)=
\int\limits_0^\infty {r'}^2dr' \, {\rho(r')
  \over r_>}
\ \  , \ 
\rho(r') = \sum_{nl}^{N-1} |R_{nl}(r')|^2 \ .
\ee
Here $r_> = \max(r,r')$ and the upper limit in the sum $N-1$ indicates
that the number of electrons in the singly ionized atomic core is
reduced by one.  The charge $Z^*_{\rm H}$ is derived from the charge density
of the occupied atomic orbitals and it neglects the exchange of the
departing photoelectron with those in the core. Thus $Z^*_{\rm H}$ provides a
convenient baseline for elucidating the exchange effects.  The charge
difference $Z^*-Z^*_{\rm H}$ is expected to be negative as
the exchange softens the atomic core and reduces its screening
capacity. In density functional theory (DFT), this effect is termed  
the exchange and correlation hole \cite{RevModPhys.61.689}.

A further model potential that we employ is that of a localized Hartree-Fock
(LHF) potential generated from a known continuous orbital calculated
in a frozen HF core \cite{0022-3700-11-24-007}.  The
radial Schr\"odinger equation with the atomic Hamiltonian \eref{Hat}
can be rewritten such that the LHF is expressed in terms of the known
HF radial orbital and its second derivative
\be
\label{LHF}
V_{\rm HF}(r) = {\kappa^2\over2}-{\ell(\ell+1)\over 2r^2}+ 
{P''_{\kappa \ell}(r) \over P_{\kappa \ell}(r)} \ .
\ee
The LHF should be weakly sensitive to the choice of the momentum
$\kappa$ and the orbital momentum $\ell$. For practical reasons, we
chose $\kappa=0.01$ and $\ell=0$ to avoid  multiple nodes of
$P_{\kappa \ell}(r)$ where the RHS of \Eref{LHF} diverges. The effective
charge $Z^*=-rV_{\rm HF}(r)$ derived from \Eref{LHF} is a smooth function
outside of these nodes and can be fitted with an analytical expression
\be
Z^*_{\rm HF}(r) = (Z_0-1)e^{-ar}+1 \ .
\label{ZHF}
\ee
This fit with $a=2.29$ for Ne and $a=2.11$ for Ar is shown on the top
panels of \Fref{Fig1} and \Fref{Fig2}, respectively. 

The $p=0$ term in \Eref{OEP} is analogous to the Muller potential
 introduced specifically for Ar \cite{Muller2002}
\be
\label{Muller}
V_{\rm M}(r)= -{1\over r} \Big[1+5.4\exp(-r)+11.6\exp(-3.682r)\Big] \ .
\ee
\citet{MILLER197716} suggested an alternative analytical expression
\be
Z^*_{\rm MD}(r)
= 1+{(Z_0-1)(1-r/R)^2\theta(R-r)\over 1+Cr+Dr^2} \ ,
\label{MD}
\ee
where $\theta(R-r)$ is the unit step function. The numerical
parameters $R$, $C$ and $D$ are chosen to match the
variation of the angular anisotropy parameter $\beta$ with energy across
the Cooper minimum (CM) known from experiment. The effective charges
generated with the potentials \eref{Muller} and \eref{MD} for Ar are
shown in \Fref{Fig2} along with those extracted from the OEP and LHF
potentials. As compared with Ne, the role of exchange is significantly
larger in Ar with the corresponding exchange hole being much greater. 
We also note that charge difference $Z^*-Z^*_{\rm H}$  in argon with the
LHF, and particularly MD, potentials is slightly positive at larger
distances.

The valence shell energies calculated with various model potentials
along with the experimental threshold energies are compiled in
\Tref{Tab1}. For the LHF potential, we also show in parentheses the
$\alpha$ parameters from \Eref{ZHF}.

\begin{table}
\caption{ The valence shell energies, in atomic units, calculated with
  various model potentials. The experimental thresholds are from
  \cite{NIST-ASD}. The LHF entries also contain the $\alpha$
  parameters from \Eref{ZHF}.
\bs
\label{Tab1}}
\begin{tabularx}{1.0\linewidth}{X|XXXXXX}
Method&Ne $2p$&Ar $3p$&Kr $4p$&Xe $5p$\\
\hline\hline\\
Expt   \cite{NIST-ASD}
      & 0.792 & 0.579 & 0.514 & 0.445\\
HF    & 0.850 & 0.591 & 0.524 & 0.457\\
OEP \cite{Sarsa2004163}  
      & 0.851 & 0.590 & 0.528 & 0.467\\
LHF   & 0.843(2.29) & 0.583(2.11) & 0.202(2.80) & 0.412(2.54)\\
Muller \cite{Muller2002}
      &       & 0.581 &       &      \\
MD    \cite{MILLER197716}
      &       & 0.423 & 0.203 &      \\
\end{tabularx}
\end{table}

\section{Results}
\label{results}

\subsection{Neon $2p$ shell}

\begin{figure}[h]

\epsfxsize=6.5cm
\epsffile{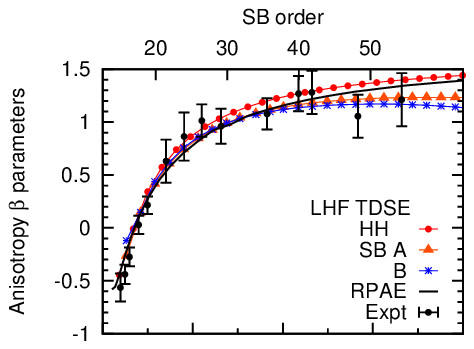}
\vs{-0.5cm}

\epsfxsize=6.5cm
\epsffile{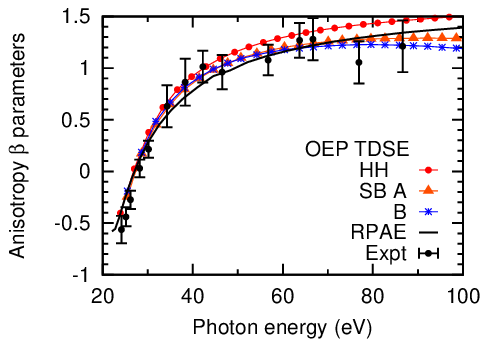}

\caption{(Color online) Angular anisotropy $\beta$ parameters for the
  Ne $2p$ valence shell extracted from the TDSE calculations with the
  LHF potential (top) and the OEP potential (bottom). The $\beta^{\rm
    HH}$ parameters extracted from the angular dependence of the high
  harmonic peaks are plotted with (red) filled circles. Same
  parameters $\beta^{\rm SB}$ extracted from the angular variation of
  the RABBITT $A$ and $B$ coefficients in \Eref{abc} are plotted with
  (orange) triangles and (blue) asterisks, respectively. The RPAE
  calculation is shown with the solid line.
The experiment \cite{0022-3700-9-5-004} is given by the points with the error bars.
\label{Fig3}} 
\end{figure}

In \Fref{Fig3} we display the angular anisotropy $\beta$ parameters
for the Ne $2p$ valence shell extracted from the TDSE calculations
with the LHF potential (top) and the OEP potential (bottom). The
$\beta^{\rm HH}$ parameters extracted from the angular dependence of
the high harmonic peaks are plotted along with the $\beta^{\rm SB}$
parameters extracted from the angular variation of the RABBITT $A$ and
$B$ parameters in \Eref{abc}. The RPAE calculation is shown with the
solid line. This calculation is known to reproduce accurately the
experimental $\beta$ parameters across the studied photon energy range
\cite{0022-3700-9-5-004}.

We see that the harmonics and sidebands TDSE calculations of $\beta$
parameters are consistent between each other and are fairly close to
the XUV-only RPAE calculation, with the LHF results marginally closer
to the RPAE than the OEP ones. In our previous work
\cite{PhysRevA.96.013408} we employed the OEP potential and quoted
$\beta^{\rm SB}\simeq0.3$ for sideband 20 (SB20) which is in
reasonable agreement with the present results of both potentials.

\begin{figure}[h]
\includegraphics[width=6cm]{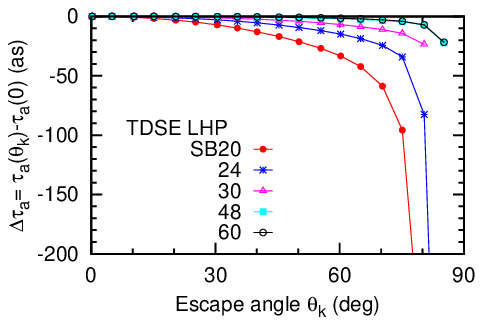}

\includegraphics[width=6cm]{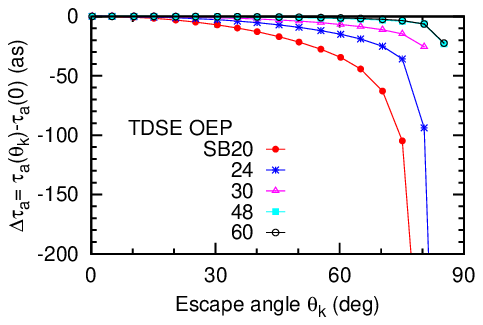}

\includegraphics[width=6cm]{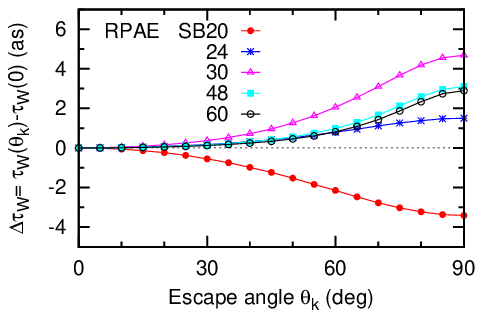}

\caption{(Color online) Angular variation of the atomic time delay
  $\Delta\tau_a = \tau_a(\theta_k)-\tau_a(0)$ in various sidebands of
  the Ne $2p$ RABBITT trace calculated with the LHF potential (top) and
the  OEP potential (middle). Bottom: angular variation of the Wigner time
  delay $\Delta\tau_{\rm W} = \tau_{\rm W}(\theta_k)-\tau_{\rm W}(0)$
  from the XUV-only RPAE calculation.
\label{Fig4}} 
\end{figure}

Angular dependence of the atomic time delay $\tau_a(\theta_k)$ as a
function of the escape angle is shown in \Fref{Fig4}. The top and
middle panels display the TDSE calculations with the LHF and OEP
potentials, respectively. The bottom panel shows the angular
dependence of the Wigner time delay $\tau_{\rm W}(\theta_k)$ from the
XUV-only RPAE calculation. We see that both TDSE calculations are
quite close to one another while the RPAE calculation suggests an
angular dependence which is an order of magnitude weaker. The
consequence being that nearly all the angular dependence of the atomic
time delay in Ne comes from the CC correction introduced by the probe
IR field. A similar observation was made in He where the Wigner time
delay is isotropic \cite{PhysRevA.94.063409}. In Ne, the Wigner time
delay is not entirely isotropic because the $2p\to\e s$ and $2p\to\e
d$ channels enter the ionization amplitude with their own spherical
harmonics, namely $Y_{00}(\theta_k)$ and $Y_{20}(\theta_k)$. However,
as a result of the Fano propensity rule \cite{PhysRevA.32.617}, the
$d$-continuum is strongly dominant and the $s$-continuum contributes
only a very weak angular modulation. We note that this situation would
change drastically near the CM in Ar and heavier noble gases where the
angular dependence of the Wigner time delay is very strong.

\begin{figure}[t]
\includegraphics[width=6cm]{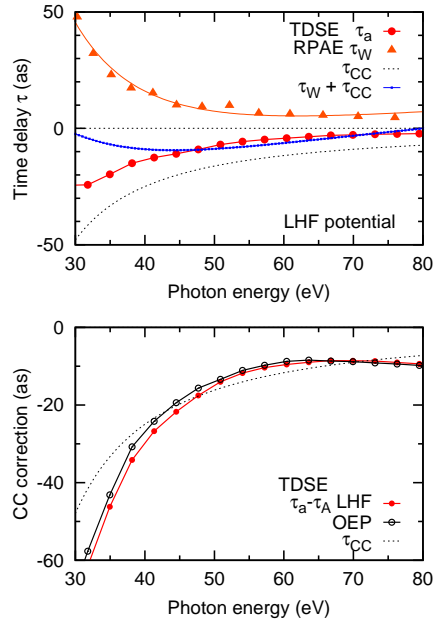}
\vs{4cm}
\caption{(Color online) Time delay in the polarization axis direction
  $\theta_k=0$. Top: the atomic time delay $\tau_a$ from the TDSE
  calculation (red filled circles) is compared with the Wigner time
  delay (orange triangle) from the RPAE calculation. The CC correction
  $\tau_{\rm CC}$ is shown with the thin dotted line whereas the sum
  $\tau_{\rm W}+ \tau_{\rm CC}$ is displayed with the (blue) dotted
  line. Bottom: the CC correction $\tau_{\rm CC}$ (thin dotted line)
  is compared with the atomic and Wigner time delay difference
  $\tau_a-\tau_{\rm W}$ from the TDSE calcualtions with the LHF and
  OEP potentials (shown with the red filled and black open circles).
\label{Fig5}} 
\end{figure}

The time delay in the polarization axis direction $\theta_k=0$ is
shown in \Fref{Fig5}. On the top panel, we compare the atomic time
delay from the TDSE calculation with the LHF potential and the Wigner
time delay $\tau_{\rm W}$ from the RPAE calculation. The hydrogenic CC
correction $\tau_{\rm CC}$, which is shown separately, is then added
to the Wigner time delay. This correction, as a function of the
photoelectron energy, is represented by the analytic expression
\be
\tau_{\rm CC}(E) = NE^{-3/2}[a\log(E)+b]
\ee
where the coefficients $N$, $a$ and $b$ are found from fitting the
regularized continuum-continuum delay shown in Fig.~7 of
\cite{Dahlstrom2012}.  We see that except for the near threshold
region where the photoelectron energy is very small and where
the regularization of $\tau_{\rm CC}$ may not be applicable, the identity
\eref{atomic} $\tau_a\simeq \tau_W+\tau_{cc}$ holds very well.

\begin{figure}[h]
\includegraphics[width=6cm]{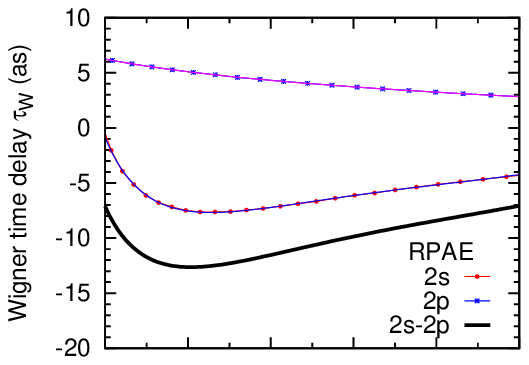}
\vs{4cm}
\caption{(Color online)  Top: the  Wigner time delay  in the  $2s$ and
  $2p$ shells of Ne and their difference. Bottom: the atomic time
  delay difference $\tau_a(2s)-\tau_a(2p)$ as measured experimentally
  by \citet{Isingerer7043} (filled circles) and
  \citet{M.Schultze06252010} (red square). The Wigner time delay
  difference $\tau_{\rm W}(2s)-\tau_{\rm W}(2p)$ (red solid line) is
  augmented by the CC correction difference $\tau_{\rm
    CC}(2s)-\tau_{\rm CC}(2p)$ (dotted line) to get the atomic time
  delay difference $\tau_a(2s)-\tau_a(2p)$ (blue solid line) which is
  compared with the calculated result of  \citet{Isingerer7043} (purple
  dashed line).
\label{Fig6}} 
\end{figure}

This utility of the hydrogenic CC correction can be used to analyze
the recent set of RABBITT measurements on Ne \cite{Isingerer7043}
where the time delay difference between the $2s$ and $2p$ shells in Ne
was determined. This analysis is shown in \Fref{Fig6}. On the top
panel, we plot the Wigner time delay from the RPAE calculation for the
individual $2s$ and $2p$ shells and their difference. On the bottom
panel, the Wigner time delay difference is augmented by that of the CC
correction. We assume that the CC correction $\tau_{\rm
  CC}$ is a universal function of the photoelectron energy and as such the CC
correction difference between shells at the same photon energy
is caused by their varying ionization potentials. The atomic time delay
difference

\ba
\label{2s2p}
\tau_a(2s)-\tau_a(2p) &=&
\tau_{\rm W}(2s)-\tau_{\rm W}(2p)
\\& +&
\tau_{\rm CC}(2s)-\tau_{\rm CC}(2p)
\nn
\ea
is compared with the RABBITT measurement and the RPA calculation
presented in \cite{Isingerer7043}. We see that both calculations
(almost indistinguishable in the scale of the figure) reproduce the
measurement \cite{Isingerer7043} very well. In contrast, the older
measurement \cite{M.Schultze06252010} deviates from the theoretical
predictions by nearly a factor of 2.

\np
\subsection{Argon $3p$ shell}

\begin{figure}[h]
\includegraphics[width=6cm]{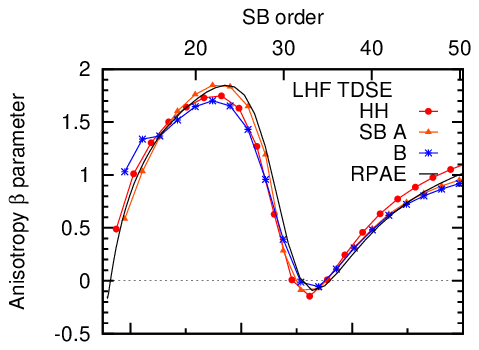}
\vs{-5mm}

\includegraphics[width=6cm]{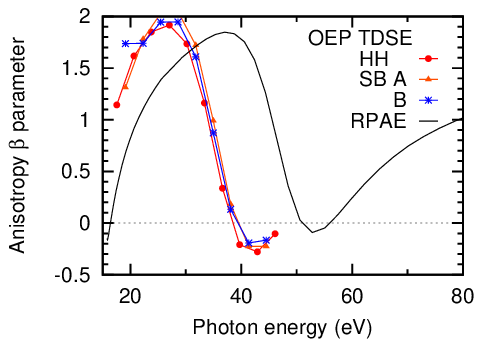}

\caption{(Color online)  
Same as \Fref{Fig3} for Ar $3p$ shell.
The experiment \cite{0022-3700-7-17-003} is given by the points with error bars.
\label{Fig7}}
\end{figure}

\begin{figure}[h]

\includegraphics[width=6cm]{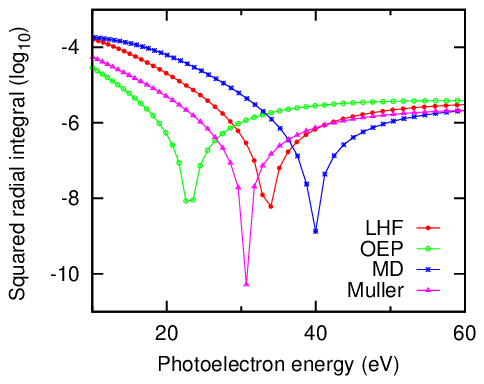}

\caption{(Color online) The squared radial integral \eref{squared}
  calculated with the LHF (red filled circles), OEP (open green
  circles), \citet{MILLER197716} (blue asterisks) and Muller \cite{Muller2002} 
   (purple triangles) potentials for Ar. The HF and RPAE results
  are shown with black dotted and solid lines respectively. 
\label{Fig8}} 
\end{figure}

The $\beta$ parameters for the Ar $3p$ shell extracted from
the angular dependence of the high harmonic peaks and sidebands are
shown in \Fref{Fig7}. The TDSE calculations performed with the LHF and
OEP potentials are shown on the top and bottom panels, respectively.
The three sets of $\beta$ parameters are compared with the RPAE
calculation and the experiment  \cite{0022-3700-7-17-003}.
We observe from this figure that all three sets of $\beta$ parameters
extracted from the TDSE calculation with the LHF potential follow
closely the RPAE prediction and agree with the experiment. At the same
time, the OEP TDSE results are displaced relative to the RPAE in the
photon energy scale by as much as 10~eV. This mismatch is a
reflection of the displacement of the CM position in the
photoionization cross-section. This position can be located very
accurately from the squared radial integral
\cite{MauritssonPRA2005}
\be
\left|
\int\limits_0^\infty P_{3p}(r)  P_{Ed}(r) \ rdr
\right|^2 \ .
\label{squared}
\ee
A plot of this integral is given in \Fref{Fig8} where the radial
orbitals of the bound and continuous states have been calculated from
the Schr\"odinger equation with Hamiltonian \eref{Hat} using the LHF,
OEP, Muller \cite{Muller2002} and \citet{MILLER197716} potentials. The
equivalent value from the HF and RPAE calculations are also shown. We
see that the CM position is misplaced for each of the potentials
except the LHF. Subsequently, in the following, we present our TDSE
results calculated with the LHF potential only.

\begin{figure}[h]
\includegraphics[width=6cm]{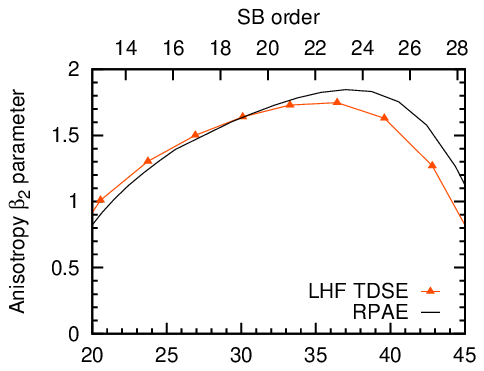}
\vs{4cm}

\caption{(Color online) Top: the anisotropy parameters derived from
  the angular variation of the high harmonic peaks.  The two sets of
  experimental values $\beta_2$ from \cite{Cirelli2018} are shown by
  open circles with error bars. The synchrotron measurement
  \cite{SOUTHWORTH1986782} is shown with black dots. The orange
  triangles connected with solid line visualize $\b^{\rm HH}$ from the
  LHF TDSE calculation whereas the black solid line displays the RPAE
  result.
Bottom: the experimental $\b_2$ and $\b_4$ parameters are shown by the
same symbols as on the top panel. The same parameters extracted from
the LHF TDSE calculation are shown with purple triangles (derived from
$A$ parameter) and blue asterisks ($B$ parameter).
\label{Fig9}}
\end{figure}

In \Fref{Fig9} we compare $\beta_2$ and $\beta_4$ parameters as
measured by \citet{Cirelli2018} and those expressed in \Eref{b24}. On the
top panel we compare $\beta_2$ and $\b^{\rm HH}$ derived from the main
harmonic peaks while on the bottom panel we display $\b_2, \b_4$ as
measured directly from the SB amplitude and as expressed via $\b^{\rm
  SB}(A)$ and $\b^{\rm SB}(B)$ in \Eref{b24}. We see that the $\b_4$
parameters compare rather favourably whereas the  $\b_2$
parameters are a bit higher than in the experiment. 

\begin{figure}[h]

\includegraphics[width=6cm]{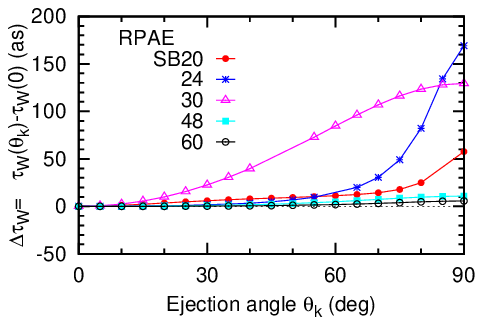}

\includegraphics[width=6cm]{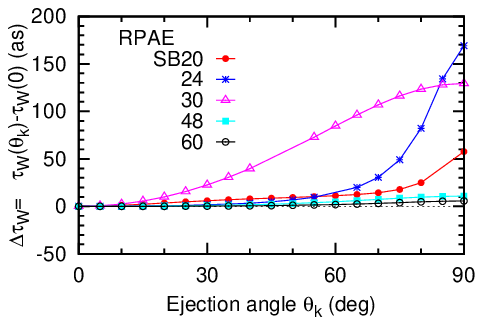}

\caption{(Color online) Top: angular variation of the atomic time
  delay $\Delta\tau_a = \tau_a(\theta_k)-\tau_a(0)$ in various
  sidebands of the Ar $3p$ RABBITT trace calculated with the LHF
  potential. Bottom: angular variation of the Wigner time delay
  $\Delta\tau_{\rm W} = \tau_{\rm W}(\theta_k)-\tau_{\rm W}(0)$ from
  the XUV-only RPAE calculation. The angular variation of time delay
  for SB32 from \cite{0953-4075-47-12-124012} is shown for comparison.
\label{Fig10}} 
\end{figure}

The angular variations of the atomic time delay $\Delta\tau_a =
\tau_a(\theta_k)-\tau_a(0)$ in various sidebands of the Ar $3p$
RABBITT trace, and the Wigner time delay angular variation
$\Delta\tau_{\rm W} = \tau_{\rm W}(\theta_k)-\tau_{\rm W}(0)$ at the
same photon energies, are displayed in \Fref{Fig10} (top and bottom
panels respectively).  In stark contrast to the analogous set of data
for Ne $2p$ shown in \Fref{Fig4}, the angular variation of the Wigner
time delay for Ar $3p$ is of the same order of magnitude, and is
almost identical for SB30 near the CM. As a reference, in both
panels of \Fref{Fig10},  the LOPT calculation \cite{0953-4075-47-12-124012}
 for SB32 is shown. Beyond the CM (SB48 and SB60), the angular variation
of the Wigner time delay flattens whereas the same variation of the
atomic time delay changes its sign and simultaneously lessens in
magnitude.

\begin{figure}[h]

\includegraphics[width=6cm]{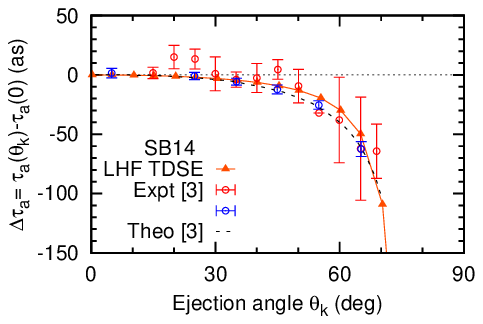}
\vs{4cm}

\caption{(Color online) Same as \Fref{Fig10} for SB14 (top) and
  SB16 (bottom). Two sets of measurements from \cite{Cirelli2018}
  are shown by open circles with error bars. The LOPT result from the
  same work is visualized by a dashed line. The LHF TDSE result is
  shown with orange triangles connected by the solid line. The bottom
  panel also shows the calculation from  \cite{Cirelli2018} which
  includes the Fano resonance (black solid line).
\label{Fig11}}
\end{figure}

In \Fref{Fig11} we compare the angular variation of the atomic time
delay $\Delta\tau_a = \tau_a(\theta_k)-\tau_a(0)$ in SB14 (top) and
SB16. In the experiment \cite{Cirelli2018}, SB16 is tuned in resonance
with the $4s^{-1}5p$ autoionizing state while SB14 is off the
resonance. For SB14 we find a fairly good agreement between the
experiment and the present LHF TDSE calculation. The LOPT calculation
reported in \cite{Cirelli2018} is also very close.  For SB16 both the
TDSE and LOPT calculations predict considerably weaker angular
dependence than in the experiment and the calculation which accounts
for resonance by the Fano configuration interaction formalism.

\begin{figure}[h]
\includegraphics[width=6cm]{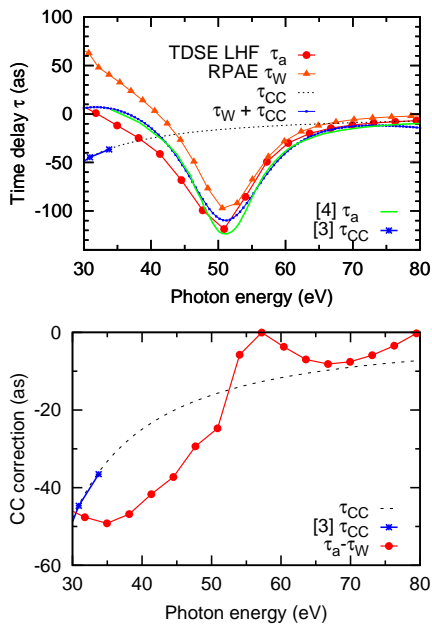}

\vs{4cm}
\caption{(Color online) 
Same as \Fref{Fig5} for Ar $3p$ shell. In addition, the atomic time
delay $\tau_a$ from the LOPT calculation
\cite{0953-4075-47-12-124012} and the CC correction
$\tau_{\rm CC} = (\phi^-_{\rm CC}-\phi^+_{\rm CC})/ 2\w$ obtained from
 the phases $\phi^{\pm}_{\rm CC}$ reported in \cite{Cirelli2018} are
shown. 
\label{Fig12}} 
\end{figure}

Various time delays for the Ar $3p$ shell in the zero angle
polarization direction are shown in \Fref{Fig12}. On the top panel, we
display the atomic time delay $\tau_a$ from the TDSE LHF calculation,
the Wigner time delay $\tau_{\rm W}$ from the RPAE calculation, the
regularized hydrogenic CC correction $\tau_{\rm CC}$ and their sum
$\tau_{\rm W}+ \tau_{\rm CC}$. We also show the atomic time delay
$\tau_a$ from the LOPT calculation \cite{0953-4075-47-12-124012}. The
latter is almost indistinguishable from the sum $\tau_{\rm
  W}+\tau_{\rm CC}$, but visibly different from the TDSE calculation
for $\tau_a$.
On the bottom panel we show the hydrogenic $\tau_{\rm CC}$ and the
argon specific value $\tau_{\rm CC} = (\phi^-_{\rm CC}-\phi^+_{\rm
  CC})/ 2\w$ obtained from the phases $\phi^{\pm}_{\rm CC}$ reported
in \cite{Cirelli2018}. Both values, which are remarkably close, are
compared with the difference $\tau_a-\tau_{\rm W}$.  Unlike in the Ne
$2p$ case, displayed on the bottom panel of \Fref{Fig5}, these two
derivations of the CC correction give quite different results. This
difference may, in principle, be attributed to the different
approximations used in TDSE-LHF and RPAE calculations. The former
employes a localized version of the HF potential and neglects the
correlation while the latter gives the full account to the exchange
and inter-shell correlation. However, the same calculations return
quite similar sets of $\beta$ parameters. As such it is more likely
that the hydrogenic approximation to $\tau_{\rm CC}$ breaks for the
argon $3p$ shell.

\begin{figure}[h]
\includegraphics[width=6cm]{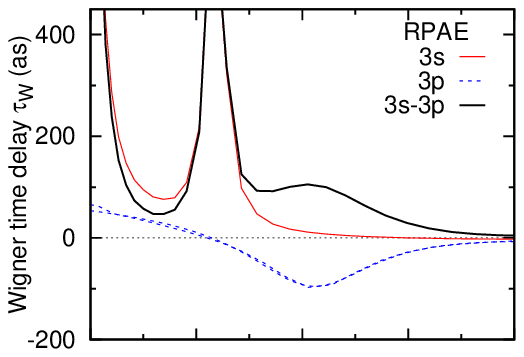}
\vs{4cm}
\caption{(Color online)  Top: the  Wigner time delay  in the  $3s$ and
  $3p$ shells of Ar and their difference. Bottom: the atomic time
  delay difference $\tau_a(3s)-\tau_a(3p)$ as measured experimentally
  by \citet{PhysRevA.85.053424} (filled circles). The Wigner time delay
  difference $\tau_{\rm W}(3s)-\tau_{\rm W}(3p)$ (red solid line) is
  augmented by the CC correction difference $\tau_{\rm
    CC}(3s)-\tau_{\rm CC}(3p)$ (dotted line) to get the atomic time
  delay difference $\tau_a(3s)-\tau_a(3p)$ (blue dashed line).
\label{Fig13}} 
\end{figure}

This break down may have implications to theoretical
interpretation of the time delay difference in the valence shell of Ar
shown in \Fref{Fig13}. Here the atomic time delay difference
\ba
\label{3s3p}
\tau_a(3s)-\tau_a(3p) &=&
\tau_{\rm W}(3s)-\tau_{\rm W}(3p)
\\& +&
\tau_{\rm CC}(3s)-\tau_{\rm CC}(3p)
\nn
\ea
is computed with the hydrogenic CC corrections and compared with the
RABBITT measurement \cite{PhysRevA.85.053424}. As $\tau_{\rm CC}(3p)$
deduced from the present TDSE calculation is more negative by about
20~as near the 40~eV mark as compared to the hydrogenic estimate, the atomic time delay
difference estimated from \Eref{3s3p} will be shifted upwards by the
same amount. It will make the disagreement with the measurement
\cite{PhysRevA.85.053424} even worse. The present TDSE calculation is not
able to give an estimate to $\tau_{\rm CC}(3s)$ as ionization of this
shell is strongly correlated with that of the valence $3p$ shell and
goes beyond the SAE approximation.

\begin{figure}[h]

\includegraphics[width=6cm]{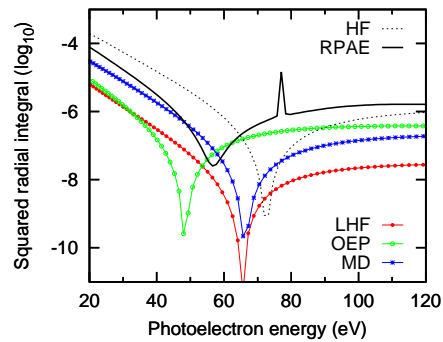}

\caption{(Color online) The squared radial integral \eref{squared}
  calculated with the LHF (red filled circles), OEP (open green
  circles), Muller \cite{Muller2002} (blue asterisks) and
  Miller\&Dow \cite{MILLER197716} (purple triangles) potentials for Kr. The HF and RPAE results
  are  shown with black dotted and solid lines respectively. 
\label{Fig14}} 
\end{figure}

\subsection{Krypton $4p$ shell}

We test validity of various effective potentials for Kr by determining
the CM position in the $4p$ photoionization cross-section. We do so by
comparing the squared radial integrals \eref{squared} calculated with
the bound state $4p$ orbital and the continuous $d$-wave obtained from
the radial Schr\"odinger equation \Eref{Hat}. This comparison is shown in
\Fref{Fig14}. Unlike in the case of Ar $3p$ photoionization,
illustrated in \Fref{Fig8}, the CM position calculated in
the HF and RPAE differs by nearly 20~eV. This is so because of the
influence of the inter-shell correlation between the $4p$ and $3d$
shells and which is accounted for in the RPAE but not in the HF
calculation. This correlation is absent in the case of Ar $3p$ as the
$3d$ shell is vacant for this atom. The CM position calculated with
the LHF and MD potentials is in between the HF and RPAE whereas the
OEP calculation displaces the CM to lower energies very
significantly. We discard the OEP in the following.

\begin{figure}[h]
\includegraphics[width=6cm]{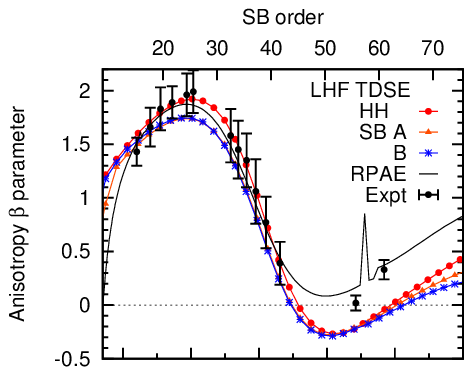}

\includegraphics[width=6cm]{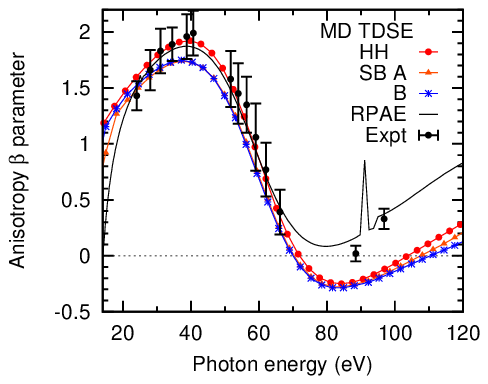}

\caption{(Color online) Same as \Fref{Fig3} for the Kr $4p$ shell. The
  experimental data are from \cite{0022-3700-10-16-015}.
\label{Fig15}}
\end{figure}

The three sets of angular anisotropy $\beta$ parameters extracted from
the high harmonic peaks and the side bands are shown in \Fref{Fig15}
calculated with the LHF (top) and MD (bottom) potentials. We see that
agreement between the TDSE and RPAE calculations is generally good but
these calculations diverge at higher photon energies. This
occurs well below the $3d$ threshold whose position can be identified
by the converging autoionization resonances visible in the RPAE curve.  The
experiment \cite{0022-3700-10-16-015} clearly favors the RPAE
calculation. Partial agreement between the TDSE calculations with the
LHF and MD potentials, the RPAE and the experiment may be
somewhat fortuitous given a strong deviation of the TDSE 
binding energies from the experimental threshold (see
\Tref{Tab1}). Should the $\beta$ parameters in \Fref{Fig15} be plotted
versus the photoelectron energy, this agreement will disappear.

\subsection{Xenon $5p$ shell}

\begin{figure}[h]

\includegraphics[width=6cm]{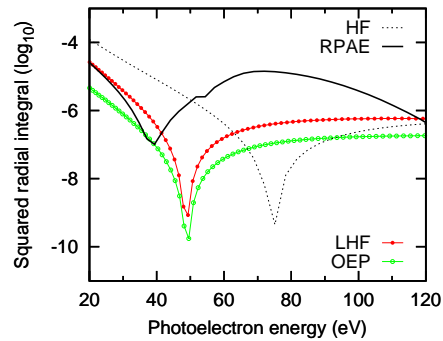}

\caption{(Color online) The squared radial integral \eref{squared}
  calculated with the LHF (red filled circles), and OEP (open green
  circles) potentials for Xe. The HF and RPAE results
  are given by the black dotted and solid lines respectively. 
\label{Fig16}} 
\end{figure}

This tendency of deviation of the TDSE calculations with various local
potentials from the RPAE and experiment is aggravated further in
Xe. As an illustration, we show in \Fref{Fig16} the CM position
deduced from the squared radial integral \eref{squared}. Firstly, we
observe that the HF and RPAE results diverge by as much as 40~eV. This
is a clear sign of a very strong correlation between the $5p$ and $4d$
shells accounted for in the RPAE but missing in the HF. Second, both
the LHF and OEP give the CM position which is displaced by 20~eV from
the RPAE for the same reason.

It is well known that missing the inter-shell correlation between the
$5p$ and $4d$ shells in Xe has a profound effect on the anisotropy
$\beta$ parameter. It becomes strongly displaced relative to the
experiment as shown graphically in Fig.~1 of \cite{West:80}. We
therefore do not expect any reasonable agreement of the presently
employed TDSE/SAE model with the experiment either.

\section{Conclusions}

We presented a series of simulations and their analysis for the
angular dependent RABBITT traces in the valence shells of noble gas
atoms from Ne to Xe. Our simulations are based on numerical solutions
of the one-electron TDSE driven with the XUV ionizing field and the IR
probing pulse. Exchange between the departing photoelectron and the
ionized atomic core is accounted for by various effective one-electron
potentials. The accuracy of this account is tested by making
comparison with the Hartree-Fock approximation which includes the
exchange by constriction.  The inter-shell correlation between the
valence $np$ and sub-valent $ns, (n-1)d$ shells are neglected in
a one-electron TDSE. To elucidate the strength of this correlation, we
compare the TDSE results with the RPAE calculation which is known to
account for the inter-shell correlation very accurately. 
However, the RPAE is unable to account for ultrafast electron
dynamics and designed for much slower ionization processes initiated
by long pulses of synchrotron radiation.

We focus our analysis on the anisotropy $\beta$ parameter which is
extracted from the angular dependence of the high harmonic peaks as
well as sideband RABBITT oscillation amplitude $A$ and $B$
factors. Within the scope of the soft photon approximation, all the
three sets of $\beta$ should be in agreement which was found to be the
case. This streamlines considerably the analysis of a angular resolved
RABBITT measurement and makes redundant the introduction of multiple
sets of angular anisotropy parameters which was made by
\citet{Cirelli2018}. The phase of the RABBITT oscillation is converted
to the angular dependent time delay which is compared with the RPAE
calculations. The time delay in the polarization direction is used to
test accuracy of the hydrogenic CC correction.

Our results can be broadly categorized into the two groups. In lighter
atoms, Ne and Ar, the single active electron  model is generally
valid. The Ne calculations are particularly robust with all the tested
effective potentials producing accurate results close to the RPAE
predictions both for the angular anisotropy and the time delay. In Ar,
because of the appearance of the Cooper minimum, the TDSE calculations
become very sensitive to the choice of the effective potential and a
simple analytic fit to the localized HF potential produces the best
results for $\beta$ parameters. At the same time, this calculation
suggests deviation of the CC correction from the regularized
hydrogenic expression. Because of the Cooper minimum, the angular
variation of the Wigner time delay is of the same magnitude as the
that variation of the atomic time delay. In Ne, the angular variation
of the Wigner time delay is negligible.

In heavier atoms, in Kr and particularly in Xe, the inter-shell
correlation between the valence $np$-shell and sub-valent $(n-1)d$
shell becomes very strong. In Kr, with some choice of effective
potentials, the present model can return sensible Cooper minimum
position and $\beta$ parameters away from the $(n-1)d$ shell
threshold. In Xe, no effective potential is expected to replace the
strong effect of inter-shell correlation and the present model is
generally invalid.

Our findings are of importance to the theoretical analysis of angular resolved
RABBITT measurements. Particularly that there is a linear
dependence of the $\b_2$ and $\b_4$ parameters which can be derived
from the single set describing the whole RABBITT measurement, both the
high harmonic peaks and the side bands.  This set can be easily
compared with predictions of the RPAE theory which is valid for all
noble gas atoms. These $\beta$ parameters can also be tested
against the XUV only measurements
\cite{0022-3700-9-5-004,0022-3700-7-17-003,0022-3700-10-16-015,
  West:80,SOUTHWORTH1986782}.

This work is a  step forward in resolving the persistent
controversy in the time delay measurement in Ar
\cite{PhysRevA.85.053424}. However, as the measurement involved
both the valence $3p$ and the sub-valent $3s$ shells, we are unable
to conclusively do so. The $3s$ shell in argon is strongly
correlated with the $3p$ shell and this inter-shell correlation goes
beyond the scope of the present model. 

The model, as it stands now, can be applied to the sub-valent Kr $3d$
and  Xe $4d$ shells which are not effected strongly by inter-shell
correlation with outer valence shells. The $nd$ correlation with inner
core is only noticeable near corresponding deeper thresholds. We can
also easily incorporate the effect of a fullerene cage
\cite{PhysRevA.96.053407} to model a RABBITT process in encapsulated
atoms.
Eventually, we  will attempt to generalize our model to account for
inter-shell correlation. This will require considerable development of
the existing one-electron TDSE code.

\section*{Acknowledgements}

We acknowledge our appreciation to Igor Ivanov for
sharing his expertise in numerical modeling of RABBITT. 
The authors are also greatly indebted to Serguei Patchkovskii who
placed his iSURF TDSE code at their disposal.
Resources of the National Computational Infrastructure facility were
 employed.

\np
\section*{Appendix: RABBITT in soft photon approximation}
\setcounter{equation}{0}
\renewcommand{\theequation}{A\arabic{equation} \rm}

We start from Eqs.~(10) and (11) of \citet{Maquet2007} and write the
amplitude of the XUV  photon  absorption 
modulated by absorption  ($+$) or emission ($-$) of~$n$~IR photons
as
\ba
S\phantom{_n} &=& \sum_{n=-\infty}^{+\infty}
S_n
\delta(k^2/2-E_0-\omega_x-n\omega)
\\
S_n&=& -2\pi i J_{-n}({\bm a}_0\cdot\k_n)
\exp[-i(\phi_x+n\phi)]
\la\k_n|{\bm \epsilon}\cdot\nabla|i\ra
\nn
\ea
with
$ k_n\simeq[2(E_0+\omega_x+n\omega)]^{1/2} $ being the shifted
momentum of the photoelectron and $\omega_x$, $\phi_x$ and $\omega$, $\phi$ 
as the XUV and IR frequencies and phase shifts respectively. 
Here the matrix element $\la\k_n|{\bm
  \epsilon}\cdot\nabla|i\ra$ of the XUV photon absorption is written
in the velocity gauge.

In RABBITT we are only interested in the $n=1$ and $n=-1$
sidebands. Their corresponding amplitudes are
\ba
S_{\pm1}
\equiv
{\cal M}^{(\pm)}_{\k_{\pm1}}
&=& -2\pi i J_{\mp1}({\bm a}_0\cdot\k_{\pm 1})
\\&\times&
\exp[\mp i\phi^{\pm}_{\rm CC}]
\la{\k_{\pm 1}}|{\bm \epsilon}\cdot\nabla|i\ra
\ .
\nn
\ea
Here we introduced  the phase $\phi^\pm_{\rm CC}$ associated with the
continuum-continuum transition in absorption or emission of
an IR photon. For simplicity we have dropped the XUV phase $\phi_x$ and thus neglected the
harmonic group delay.
Using the transformation  $J_{-n} = (-1)^nJ_n$ we write
\ba
S_{+1}+S_{-1} &=&  \\
&&\hs{-2cm}-2\pi i \Big[
J_1({\bm a}_0\cdot\k_{-1})
\la{\k_{-1}}|{\bm \epsilon}\cdot\nabla|i\ra \
e^{\ds-i\phi^+_{\rm CC}}
\nn\\&&\hs{-1.3cm}-
J_1({\bm a}_0\cdot\k_{+1})
\la{\k_{+1}}|{\bm \epsilon}\cdot\nabla|i\ra \
e^{\ds+i\phi^-_{\rm CC}}
\Big]
\ .
\nn
\ea
We can relate the phases of the dipole matrix elements with the soft
photon shifted momenta by the phase energy derivative,
\ba
\nn
\la{\k_{\pm1}}|{\bm \epsilon}\cdot\nabla|i\ra
&\approx&
\la\k|{\bm \epsilon}\cdot\nabla|i\ra
e^{\pm i\omega\partial\delta_{m_i}(\k)/\partial E}
\\
\delta_{m_i}(\k) &=& \arg \la\k|{\bm \epsilon}\cdot\nabla|i\ra
\ .
\ea
Further, we assume
$
J_1({\bm \alpha}_0\cdot\k_{\pm1})\approx
J_1({\bm \alpha}_0\cdot\k)
$
and subsequently find the magnitude of the RABBITT signal \eref{abc} to be proportional to
\ba
\nn
{\rm Re} \left[
{\cal M}^{(-)}_{\k}
{\cal M}^{*(+)}_{\k}
\right]
&\propto& 
\big|J_1({\bm \alpha}_0\cdot\k)\big|^2
\Big[
\sum_{m_i}
|\la\chi_\k|{\bm \epsilon}\cdot\nabla|\psi_i\ra|^2
\Big]
\\
&\propto& 
\cos^2\theta\Big[1+\beta P_2(\cos\theta)\Big]
\ .
\ea
Here we used the expansion 
$
J_1(x)\simeq x/2 + O(x^3)
$
valid for a weak IR field and accordingly small parameter
${\bm \alpha}_0={\bm F}_0/\omega^2
\ .
$
We also performed the angular momentum projection summation \cite{A90}  
\be
\sum_{m_i}
|\la\k|{\bm \epsilon}\cdot\nabla|i\ra|^2
\propto {\sigma_i\over 4\pi}
\Big[1+\beta P_2(\cos\theta)\Big]
\ ,
\ee
where $\beta$ is the angular anisotropy parameter  and
$\sigma_i$ is the photoionization cross-section of the $i$-th atomic shell.

The atomic time delay is given by
\ba
\tau
&=& 
{1\over 2\w}
\arg
\left[
{\cal M}^{(-)}_{\k}
{\cal M}^{*(+)}_{\k}
\right]
\\&\equiv&
{1\over 2\w}
\arg\Big[
\sum_{m_i}
|c_{m_i}|^2
e^{2i\phi_{m_i}}
\Big]
+{\phi^-_{\rm CC}-\phi^+_{\rm CC}\over 2\w}  
\nn\\
&\equiv& \tau_{\rm W} + \tau_{\rm CC}
\nn
\ea
where we have used the shorthand
$$
\phi_{m_i} = \w \ {\partial \delta_{m_i}(\k)\over\partial E}
\ , \ 
c_{m_i} = \la\chi_\k|{\bm \epsilon}\cdot\nabla|\psi_i\ra
$$
for the quantities associated with the XUV photon absorption which
define the Wigner time delay $\tau_{\rm W}$.


\begin{thebibliography}{36}
\expandafter\ifx\csname natexlab\endcsname\relax\def\natexlab#1{#1}\fi
\expandafter\ifx\csname bibnamefont\endcsname\relax
  \def\bibnamefont#1{#1}\fi
\expandafter\ifx\csname bibfnamefont\endcsname\relax
  \def\bibfnamefont#1{#1}\fi
\expandafter\ifx\csname citenamefont\endcsname\relax
  \def\citenamefont#1{#1}\fi
\expandafter\ifx\csname url\endcsname\relax
  \def\url#1{\texttt{#1}}\fi
\expandafter\ifx\csname urlprefix\endcsname\relax\def\urlprefix{URL }\fi
\providecommand{\bibinfo}[2]{#2}
\providecommand{\eprint}[2][]{\url{#2}}

\bibitem[{\citenamefont{Laurent et~al.}(2012)\citenamefont{Laurent, Cao, Li,
  Wang, Ben-Itzhak, and Cocke}}]{PhysRevLett.109.083001}
\bibinfo{author}{\bibfnamefont{G.}~\bibnamefont{Laurent}},
  \bibinfo{author}{\bibfnamefont{W.}~\bibnamefont{Cao}},
  \bibinfo{author}{\bibfnamefont{H.}~\bibnamefont{Li}},
  \bibinfo{author}{\bibfnamefont{Z.}~\bibnamefont{Wang}},
  \bibinfo{author}{\bibfnamefont{I.}~\bibnamefont{Ben-Itzhak}},
  \bibnamefont{and} \bibinfo{author}{\bibfnamefont{C.~L.} \bibnamefont{Cocke}},
  \emph{\bibinfo{title}{Attosecond control of orbital parity mix interferences
  and the relative phase of even and odd harmonics in an attosecond pulse
  train}}, \bibinfo{journal}{Phys. Rev. Lett.} \textbf{\bibinfo{volume}{109}},
  \bibinfo{pages}{083001} (\bibinfo{year}{2012}).

\bibitem[{\citenamefont{Heuser et~al.}(2016)\citenamefont{Heuser,
  Jim\'enez~Gal\'an, Cirelli, Marante, Sabbar, Boge, Lucchini, Gallmann,
  Ivanov, Kheifets et~al.}}]{PhysRevA.94.063409}
\bibinfo{author}{\bibfnamefont{S.}~\bibnamefont{Heuser}},
  \bibinfo{author}{\bibfnamefont{A.}~\bibnamefont{Jim\'enez~Gal\'an}},
  \bibinfo{author}{\bibfnamefont{C.}~\bibnamefont{Cirelli}},
  \bibinfo{author}{\bibfnamefont{C.}~\bibnamefont{Marante}},
  \bibinfo{author}{\bibfnamefont{M.}~\bibnamefont{Sabbar}},
  \bibinfo{author}{\bibfnamefont{R.}~\bibnamefont{Boge}},
  \bibinfo{author}{\bibfnamefont{M.}~\bibnamefont{Lucchini}},
  \bibinfo{author}{\bibfnamefont{L.}~\bibnamefont{Gallmann}},
  \bibinfo{author}{\bibfnamefont{I.}~\bibnamefont{Ivanov}},
  \bibinfo{author}{\bibfnamefont{A.~S.} \bibnamefont{Kheifets}},
  \bibnamefont{et~al.}, \emph{\bibinfo{title}{Angular dependence of
  photoemission time delay in helium}}, \bibinfo{journal}{Phys. Rev. A}
  \textbf{\bibinfo{volume}{94}}, \bibinfo{pages}{063409}
  (\bibinfo{year}{2016}).

\bibitem[{\citenamefont{Cirelli et~al.}(2018)\citenamefont{Cirelli, Marante,
  Heuser, Petersson, Gal\'an, Argenti, Zhong, Busto, Isinger, Nandi
  et~al.}}]{Cirelli2018}
\bibinfo{author}{\bibfnamefont{C.}~\bibnamefont{Cirelli}},
  \bibinfo{author}{\bibfnamefont{C.}~\bibnamefont{Marante}},
  \bibinfo{author}{\bibfnamefont{S.}~\bibnamefont{Heuser}},
  \bibinfo{author}{\bibfnamefont{C.~L.~M.} \bibnamefont{Petersson}},
  \bibinfo{author}{\bibfnamefont{A.~J.} \bibnamefont{Gal\'an}},
  \bibinfo{author}{\bibfnamefont{L.}~\bibnamefont{Argenti}},
  \bibinfo{author}{\bibfnamefont{S.}~\bibnamefont{Zhong}},
  \bibinfo{author}{\bibfnamefont{D.}~\bibnamefont{Busto}},
  \bibinfo{author}{\bibfnamefont{M.}~\bibnamefont{Isinger}},
  \bibinfo{author}{\bibfnamefont{S.}~\bibnamefont{Nandi}},
  \bibnamefont{et~al.}, \emph{\bibinfo{title}{Anisotropic photoemission time
  delays close to a {Fano} resonance}}, \bibinfo{journal}{Nature Comm.}
  \textbf{\bibinfo{volume}{9}}, \bibinfo{pages}{955} (\bibinfo{year}{2018}).

\bibitem[{\citenamefont{Dahlström and
  Lindroth}(2014)}]{0953-4075-47-12-124012}
\bibinfo{author}{\bibfnamefont{J.~M.} \bibnamefont{Dahlström}}
  \bibnamefont{and} \bibinfo{author}{\bibfnamefont{E.}~\bibnamefont{Lindroth}},
  \emph{\bibinfo{title}{Study of attosecond delays using perturbation diagrams
  and exterior complex scaling}}, \bibinfo{journal}{J. Phys. B}
  \textbf{\bibinfo{volume}{47}}(\bibinfo{number}{12}), \bibinfo{pages}{124012}
  (\bibinfo{year}{2014}).

\bibitem[{\citenamefont{Hockett}(2017)}]{0953-4075-50-15-154002}
\bibinfo{author}{\bibfnamefont{P.}~\bibnamefont{Hockett}},
  \emph{\bibinfo{title}{Angle-resolved {RABBITT}: theory and numerics}},
  \bibinfo{journal}{J. Phys. B}
  \textbf{\bibinfo{volume}{50}}(\bibinfo{number}{15}), \bibinfo{pages}{154002}
  (\bibinfo{year}{2017}).

\bibitem[{\citenamefont{Ivanov and Kheifets}(2017)}]{PhysRevA.96.013408}
\bibinfo{author}{\bibfnamefont{I.~A.} \bibnamefont{Ivanov}} \bibnamefont{and}
  \bibinfo{author}{\bibfnamefont{A.~S.} \bibnamefont{Kheifets}},
  \emph{\bibinfo{title}{Angle-dependent time delay in two-color {XUV+IR}
  photoemission of {He} and {Ne}}}, \bibinfo{journal}{Phys. Rev. A}
  \textbf{\bibinfo{volume}{96}}, \bibinfo{pages}{013408}
  (\bibinfo{year}{2017}).

\bibitem[{\citenamefont{Muller}(2002{\natexlab{a}})}]{MullerAPB2002}
\bibinfo{author}{\bibfnamefont{H.}~\bibnamefont{Muller}},
  \emph{\bibinfo{title}{Reconstruction of attosecond harmonic beating by
  interference of two-photon transitions}}, \bibinfo{journal}{Appl. Phys. B}
  \textbf{\bibinfo{volume}{74}}, \bibinfo{pages}{s17}
  (\bibinfo{year}{2002}{\natexlab{a}}).

\bibitem[{\citenamefont{Toma and Muller}(2002)}]{TomaJPB2002}
\bibinfo{author}{\bibfnamefont{E.~S.} \bibnamefont{Toma}} \bibnamefont{and}
  \bibinfo{author}{\bibfnamefont{H.~G.} \bibnamefont{Muller}},
  \emph{\bibinfo{title}{Calculation of matrix elements for mixed
  extreme-ultraviolet–infrared two-photon above-threshold ionization of
  argon}}, \bibinfo{journal}{J. Phys. B}
  \textbf{\bibinfo{volume}{35}}(\bibinfo{number}{16}), \bibinfo{pages}{3435}
  (\bibinfo{year}{2002}).

\bibitem[{\citenamefont{Sarsa et~al.}(2004)\citenamefont{Sarsa, G\'{a}lvez, and
  Buendia}}]{Sarsa2004163}
\bibinfo{author}{\bibfnamefont{A.}~\bibnamefont{Sarsa}},
  \bibinfo{author}{\bibfnamefont{F.~J.} \bibnamefont{G\'{a}lvez}},
  \bibnamefont{and} \bibinfo{author}{\bibfnamefont{E.}~\bibnamefont{Buendia}},
  \emph{\bibinfo{title}{Parameterized optimized effective potential for the
  ground state of the atoms {He} through {Xe}}}, \bibinfo{journal}{Atomic Data
  and Nuclear Data Tables} \textbf{\bibinfo{volume}{88}}(\bibinfo{number}{1}),
  \bibinfo{pages}{163 } (\bibinfo{year}{2004}).

\bibitem[{\citenamefont{Gu\'enot et~al.}(2012)\citenamefont{Gu\'enot,
  Kl\"under, Arnold, Kroon, Dahlstr\"om, Miranda, Fordell, Gisselbrecht,
  Johnsson, Mauritsson et~al.}}]{PhysRevA.85.053424}
\bibinfo{author}{\bibfnamefont{D.}~\bibnamefont{Gu\'enot}},
  \bibinfo{author}{\bibfnamefont{K.}~\bibnamefont{Kl\"under}},
  \bibinfo{author}{\bibfnamefont{C.~L.} \bibnamefont{Arnold}},
  \bibinfo{author}{\bibfnamefont{D.}~\bibnamefont{Kroon}},
  \bibinfo{author}{\bibfnamefont{J.~M.} \bibnamefont{Dahlstr\"om}},
  \bibinfo{author}{\bibfnamefont{M.}~\bibnamefont{Miranda}},
  \bibinfo{author}{\bibfnamefont{T.}~\bibnamefont{Fordell}},
  \bibinfo{author}{\bibfnamefont{M.}~\bibnamefont{Gisselbrecht}},
  \bibinfo{author}{\bibfnamefont{P.}~\bibnamefont{Johnsson}},
  \bibinfo{author}{\bibfnamefont{J.}~\bibnamefont{Mauritsson}},
  \bibnamefont{et~al.}, \emph{\bibinfo{title}{Photoemission-time-delay
  measurements and calculations close to the 3$s$-ionization-cross-section
  minimum in {Ar}}}, \bibinfo{journal}{Phys. Rev. A}
  \textbf{\bibinfo{volume}{85}}, \bibinfo{pages}{053424}
  (\bibinfo{year}{2012}).

\bibitem[{\citenamefont{Isinger et~al.}(2017)\citenamefont{Isinger, Squibb,
  Busto, Zhong, Harth, Kroon, Nandi, Arnold, Miranda, Dahlstr{\"o}m
  et~al.}}]{Isingerer7043}
\bibinfo{author}{\bibfnamefont{M.}~\bibnamefont{Isinger}},
  \bibinfo{author}{\bibfnamefont{R.}~\bibnamefont{Squibb}},
  \bibinfo{author}{\bibfnamefont{D.}~\bibnamefont{Busto}},
  \bibinfo{author}{\bibfnamefont{S.}~\bibnamefont{Zhong}},
  \bibinfo{author}{\bibfnamefont{A.}~\bibnamefont{Harth}},
  \bibinfo{author}{\bibfnamefont{D.}~\bibnamefont{Kroon}},
  \bibinfo{author}{\bibfnamefont{S.}~\bibnamefont{Nandi}},
  \bibinfo{author}{\bibfnamefont{C.~L.} \bibnamefont{Arnold}},
  \bibinfo{author}{\bibfnamefont{M.}~\bibnamefont{Miranda}},
  \bibinfo{author}{\bibfnamefont{J.~M.} \bibnamefont{Dahlstr{\"o}m}},
  \bibnamefont{et~al.}, \emph{\bibinfo{title}{Photoionization in the time and
  frequency domain}}, \bibinfo{journal}{Science}  (\bibinfo{year}{2017}).

\bibitem[{\citenamefont{{Schultze {\em et~al}}}(2010)}]{M.Schultze06252010}
\bibinfo{author}{\bibfnamefont{M.}~\bibnamefont{{Schultze {\em et~al}}}},
  \emph{\bibinfo{title}{{Delay in Photoemission}}}, \bibinfo{journal}{Science}
  \textbf{\bibinfo{volume}{328}}(\bibinfo{number}{5986}), \bibinfo{pages}{1658}
  (\bibinfo{year}{2010}).

\bibitem[{\citenamefont{Kheifets and Ivanov}(2010)}]{PhysRevLett.105.233002}
\bibinfo{author}{\bibfnamefont{A.~S.} \bibnamefont{Kheifets}} \bibnamefont{and}
  \bibinfo{author}{\bibfnamefont{I.~A.} \bibnamefont{Ivanov}},
  \emph{\bibinfo{title}{Delay in atomic photoionization}},
  \bibinfo{journal}{Phys. Rev. Lett.}
  \textbf{\bibinfo{volume}{105}}(\bibinfo{number}{23}), \bibinfo{pages}{233002}
  (\bibinfo{year}{2010}).

\bibitem[{\citenamefont{Moore et~al.}(2011)\citenamefont{Moore, Lysaght,
  Parker, van~der Hart, and Taylor}}]{PhysRevA.84.061404}
\bibinfo{author}{\bibfnamefont{L.~R.} \bibnamefont{Moore}},
  \bibinfo{author}{\bibfnamefont{M.~A.} \bibnamefont{Lysaght}},
  \bibinfo{author}{\bibfnamefont{J.~S.} \bibnamefont{Parker}},
  \bibinfo{author}{\bibfnamefont{H.~W.} \bibnamefont{van~der Hart}},
  \bibnamefont{and} \bibinfo{author}{\bibfnamefont{K.~T.}
  \bibnamefont{Taylor}}, \emph{\bibinfo{title}{Time delay between photoemission
  from the $2p$ and $2s$ subshells of neon}}, \bibinfo{journal}{Phys. Rev. A}
  \textbf{\bibinfo{volume}{84}}, \bibinfo{pages}{061404}
  (\bibinfo{year}{2011}).

\bibitem[{\citenamefont{Dahlstr\"om
  et~al.}(2012{\natexlab{a}})\citenamefont{Dahlstr\"om, Carette, and
  Lindroth}}]{PhysRevA.86.061402}
\bibinfo{author}{\bibfnamefont{J.~M.} \bibnamefont{Dahlstr\"om}},
  \bibinfo{author}{\bibfnamefont{T.}~\bibnamefont{Carette}}, \bibnamefont{and}
  \bibinfo{author}{\bibfnamefont{E.}~\bibnamefont{Lindroth}},
  \emph{\bibinfo{title}{Diagrammatic approach to attosecond delays in
  photoionization}}, \bibinfo{journal}{Phys. Rev. A}
  \textbf{\bibinfo{volume}{86}}, \bibinfo{pages}{061402}
  (\bibinfo{year}{2012}{\natexlab{a}}).

\bibitem[{\citenamefont{Kheifets}(2013)}]{PhysRevA.87.063404}
\bibinfo{author}{\bibfnamefont{A.~S.} \bibnamefont{Kheifets}},
  \emph{\bibinfo{title}{Time delay in valence-shell photoionization of
  noble-gas atoms}}, \bibinfo{journal}{Phys. Rev. A}
  \textbf{\bibinfo{volume}{87}}, \bibinfo{pages}{063404}
  (\bibinfo{year}{2013}).

\bibitem[{\citenamefont{Feist et~al.}(2014)\citenamefont{Feist, Zatsarinny,
  Nagele, Pazourek, Burgd\"orfer, Guan, Bartschat, and
  Schneider}}]{PhysRevA.89.033417}
\bibinfo{author}{\bibfnamefont{J.}~\bibnamefont{Feist}},
  \bibinfo{author}{\bibfnamefont{O.}~\bibnamefont{Zatsarinny}},
  \bibinfo{author}{\bibfnamefont{S.}~\bibnamefont{Nagele}},
  \bibinfo{author}{\bibfnamefont{R.}~\bibnamefont{Pazourek}},
  \bibinfo{author}{\bibfnamefont{J.}~\bibnamefont{Burgd\"orfer}},
  \bibinfo{author}{\bibfnamefont{X.}~\bibnamefont{Guan}},
  \bibinfo{author}{\bibfnamefont{K.}~\bibnamefont{Bartschat}},
  \bibnamefont{and} \bibinfo{author}{\bibfnamefont{B.~I.}
  \bibnamefont{Schneider}}, \emph{\bibinfo{title}{Time delays for attosecond
  streaking in photoionization of neon}}, \bibinfo{journal}{Phys. Rev. A}
  \textbf{\bibinfo{volume}{89}}, \bibinfo{pages}{033417}
  (\bibinfo{year}{2014}).

\bibitem[{\citenamefont{Omiste and Madsen}(2018)}]{PhysRevA.97.013422}
\bibinfo{author}{\bibfnamefont{J.~J.} \bibnamefont{Omiste}} \bibnamefont{and}
  \bibinfo{author}{\bibfnamefont{L.~B.} \bibnamefont{Madsen}},
  \emph{\bibinfo{title}{Attosecond photoionization dynamics in neon}},
  \bibinfo{journal}{Phys. Rev. A} \textbf{\bibinfo{volume}{97}},
  \bibinfo{pages}{013422} (\bibinfo{year}{2018}).

\bibitem[{\citenamefont{Morales et~al.}(2016)\citenamefont{Morales, Bredtmann,
  and Patchkovskii}}]{0953-4075-49-24-245001}
\bibinfo{author}{\bibfnamefont{F.}~\bibnamefont{Morales}},
  \bibinfo{author}{\bibfnamefont{T.}~\bibnamefont{Bredtmann}},
  \bibnamefont{and}
  \bibinfo{author}{\bibfnamefont{S.}~\bibnamefont{Patchkovskii}},
  \emph{\bibinfo{title}{isurf: a family of infinite-time surface flux
  methods}}, \bibinfo{journal}{J. Phys. B}
  \textbf{\bibinfo{volume}{49}}(\bibinfo{number}{24}), \bibinfo{pages}{245001}
  (\bibinfo{year}{2016}).

\bibitem[{\citenamefont{Maquet and Ta\"{\i}eb}(2007)}]{Maquet2007}
\bibinfo{author}{\bibfnamefont{A.}~\bibnamefont{Maquet}} \bibnamefont{and}
  \bibinfo{author}{\bibfnamefont{R.}~\bibnamefont{Ta\"{\i}eb}},
  \emph{\bibinfo{title}{Two-colour ir+xuv spectroscopies: the soft-photon
  approximation}}, \bibinfo{journal}{J. Modern Optics}
  \textbf{\bibinfo{volume}{54}}(\bibinfo{number}{13-15}), \bibinfo{pages}{1847}
  (\bibinfo{year}{2007}).

\bibitem[{\citenamefont{{Kheifets {\em et al}}}(2016)}]{PhysRevA.94.013423}
\bibinfo{author}{\bibfnamefont{A.~S.} \bibnamefont{{Kheifets {\em et al}}}},
  \emph{\bibinfo{title}{Relativistic calculations of angle-dependent
  photoemission time delay}}, \bibinfo{journal}{Phys. Rev. A}
  \textbf{\bibinfo{volume}{94}}, \bibinfo{pages}{013423}
  (\bibinfo{year}{2016}).

\bibitem[{\citenamefont{Dahlstr\"om
  et~al.}(2012{\natexlab{b}})\citenamefont{Dahlstr\"om, Gu\'enot, Kl\"under,
  Gisselbrecht, Mauritsson, Huillier, Maquet, and Ta\"ieb}}]{Dahlstrom2012}
\bibinfo{author}{\bibfnamefont{J.}~\bibnamefont{Dahlstr\"om}},
  \bibinfo{author}{\bibfnamefont{D.}~\bibnamefont{Gu\'enot}},
  \bibinfo{author}{\bibfnamefont{K.}~\bibnamefont{Kl\"under}},
  \bibinfo{author}{\bibfnamefont{M.}~\bibnamefont{Gisselbrecht}},
  \bibinfo{author}{\bibfnamefont{J.}~\bibnamefont{Mauritsson}},
  \bibinfo{author}{\bibfnamefont{A.~L.} \bibnamefont{Huillier}},
  \bibinfo{author}{\bibfnamefont{A.}~\bibnamefont{Maquet}}, \bibnamefont{and}
  \bibinfo{author}{\bibfnamefont{R.}~\bibnamefont{Ta\"ieb}},
  \emph{\bibinfo{title}{Theory of attosecond delays in laser-assisted
  photoionization}}, \bibinfo{journal}{Chem. Phys.}
  \textbf{\bibinfo{volume}{414}}, \bibinfo{pages}{53}
  (\bibinfo{year}{2012}{\natexlab{b}}).

\bibitem[{\citenamefont{Slater}(1951)}]{PhysRev.81.385}
\bibinfo{author}{\bibfnamefont{J.~C.} \bibnamefont{Slater}},
  \emph{\bibinfo{title}{A simplification of the Hartree-Fock method}},
  \bibinfo{journal}{Phys. Rev.} \textbf{\bibinfo{volume}{81}},
  \bibinfo{pages}{385} (\bibinfo{year}{1951}).

\bibitem[{\citenamefont{Miller and Dow}(1977)}]{MILLER197716}
\bibinfo{author}{\bibfnamefont{D.~L.} \bibnamefont{Miller}} \bibnamefont{and}
  \bibinfo{author}{\bibfnamefont{J.~D.} \bibnamefont{Dow}},
  \emph{\bibinfo{title}{Atomic pseudopotentials for soft x-ray excitations}},
  \bibinfo{journal}{Physics Letters A}
  \textbf{\bibinfo{volume}{60}}(\bibinfo{number}{1}), \bibinfo{pages}{16 }
  (\bibinfo{year}{1977}).

\bibitem[{\citenamefont{Muller}(2002{\natexlab{b}})}]{Muller2002}
\bibinfo{author}{\bibfnamefont{H.}~\bibnamefont{Muller}},
  \emph{\bibinfo{title}{Reconstruction of attosecond harmonic beating by
  interference of two-photon transitions}}, \bibinfo{journal}{Applied Physics
  B} \textbf{\bibinfo{volume}{74}}(\bibinfo{number}{1}), \bibinfo{pages}{s17}
  (\bibinfo{year}{2002}{\natexlab{b}}).

\bibitem[{\citenamefont{Jones and Gunnarsson}(1989)}]{RevModPhys.61.689}
\bibinfo{author}{\bibfnamefont{R.~O.} \bibnamefont{Jones}} \bibnamefont{and}
  \bibinfo{author}{\bibfnamefont{O.}~\bibnamefont{Gunnarsson}},
  \emph{\bibinfo{title}{The density functional formalism, its applications and
  prospects}}, \bibinfo{journal}{Rev. Mod. Phys.}
  \textbf{\bibinfo{volume}{61}}, \bibinfo{pages}{689} (\bibinfo{year}{1989}).

\bibitem[{\citenamefont{Wendin and Starace}(1978)}]{0022-3700-11-24-007}
\bibinfo{author}{\bibfnamefont{G.}~\bibnamefont{Wendin}} \bibnamefont{and}
  \bibinfo{author}{\bibfnamefont{A.~F.} \bibnamefont{Starace}},
  \emph{\bibinfo{title}{Perturbation theory in a strong-interaction regime with
  application to 4d-subshell spectra of ba and la}}, \bibinfo{journal}{J. Phys.
  B} \textbf{\bibinfo{volume}{11}}(\bibinfo{number}{24}), \bibinfo{pages}{4119}
  (\bibinfo{year}{1978}).

\bibitem[{\citenamefont{Ralchenko et~al.}(2011)\citenamefont{Ralchenko,
  Kramida, Reader, and {NIST ASD Team}}}]{NIST-ASD}
\bibinfo{author}{\bibfnamefont{Y.}~\bibnamefont{Ralchenko}},
  \bibinfo{author}{\bibfnamefont{A.~E.} \bibnamefont{Kramida}},
  \bibinfo{author}{\bibfnamefont{J.}~\bibnamefont{Reader}}, \bibnamefont{and}
  \bibinfo{author}{\bibnamefont{{NIST ASD Team}}}, \emph{\bibinfo{title}{{NIST}
  atomic spectra database (version 3.1.5)}} (\bibinfo{year}{2011}),
  \urlprefix\url{http://physics.nist.gov/asd}.

\bibitem[{\citenamefont{Codling et~al.}(1976)\citenamefont{Codling, Houlgate,
  West, and Woodruff}}]{0022-3700-9-5-004}
\bibinfo{author}{\bibfnamefont{K.}~\bibnamefont{Codling}},
  \bibinfo{author}{\bibfnamefont{R.~G.} \bibnamefont{Houlgate}},
  \bibinfo{author}{\bibfnamefont{J.~B.} \bibnamefont{West}}, \bibnamefont{and}
  \bibinfo{author}{\bibfnamefont{P.~R.} \bibnamefont{Woodruff}},
  \emph{\bibinfo{title}{Angular distribution and photoionization measurements
  on the 2p and 2s electrons in neon}}, \bibinfo{journal}{J. Phys. B}
  \textbf{\bibinfo{volume}{9}}(\bibinfo{number}{5}), \bibinfo{pages}{L83}
  (\bibinfo{year}{1976}).

\bibitem[{\citenamefont{Fano}(1985)}]{PhysRevA.32.617}
\bibinfo{author}{\bibfnamefont{U.}~\bibnamefont{Fano}},
  \emph{\bibinfo{title}{Propensity rules: An analytical approach}},
  \bibinfo{journal}{Phys. Rev. A} \textbf{\bibinfo{volume}{32}},
  \bibinfo{pages}{617} (\bibinfo{year}{1985}).

\bibitem[{\citenamefont{Houlgate et~al.}(1974)\citenamefont{Houlgate, Codling,
  Marr, and West}}]{0022-3700-7-17-003}
\bibinfo{author}{\bibfnamefont{R.~G.} \bibnamefont{Houlgate}},
  \bibinfo{author}{\bibfnamefont{K.}~\bibnamefont{Codling}},
  \bibinfo{author}{\bibfnamefont{G.~V.} \bibnamefont{Marr}}, \bibnamefont{and}
  \bibinfo{author}{\bibfnamefont{J.~B.} \bibnamefont{West}},
  \emph{\bibinfo{title}{Angular distribution and photoionization cross section
  measurements on the 3p and 3s subshells of argon}}, \bibinfo{journal}{J.
  Phys. B} \textbf{\bibinfo{volume}{7}}(\bibinfo{number}{17}),
  \bibinfo{pages}{L470} (\bibinfo{year}{1974}).

\bibitem[{\citenamefont{Mauritsson et~al.}(2005)\citenamefont{Mauritsson,
  Gaarde, and Schafer}}]{MauritssonPRA2005}
\bibinfo{author}{\bibfnamefont{J.}~\bibnamefont{Mauritsson}},
  \bibinfo{author}{\bibfnamefont{M.~B.} \bibnamefont{Gaarde}},
  \bibnamefont{and} \bibinfo{author}{\bibfnamefont{K.~J.}
  \bibnamefont{Schafer}}, \emph{\bibinfo{title}{Accessing properties of
  electron wave packets generated by attosecond pulse trains through
  time-dependent calculations}}, \bibinfo{journal}{Phys. Rev. A}
  \textbf{\bibinfo{volume}{72}}, \bibinfo{pages}{013401}
  (\bibinfo{year}{2005}).

\bibitem[{\citenamefont{Southworth et~al.}(1986)\citenamefont{Southworth, Parr,
  Hardis, Dehmer, and Holland}}]{SOUTHWORTH1986782}
\bibinfo{author}{\bibfnamefont{S.}~\bibnamefont{Southworth}},
  \bibinfo{author}{\bibfnamefont{A.}~\bibnamefont{Parr}},
  \bibinfo{author}{\bibfnamefont{J.}~\bibnamefont{Hardis}},
  \bibinfo{author}{\bibfnamefont{J.}~\bibnamefont{Dehmer}}, \bibnamefont{and}
  \bibinfo{author}{\bibfnamefont{D.}~\bibnamefont{Holland}},
  \emph{\bibinfo{title}{Calibration of a monochromator/spectrometer system for
  the measurement of photoelectron angular distributions and branching
  ratios}}, \bibinfo{journal}{Nucl. Instr. Methods A}
  \textbf{\bibinfo{volume}{246}}(\bibinfo{number}{1}), \bibinfo{pages}{782 }
  (\bibinfo{year}{1986}).

\bibitem[{\citenamefont{Miller et~al.}(1977)\citenamefont{Miller, Dow,
  Houlgate, Marr, and West}}]{0022-3700-10-16-015}
\bibinfo{author}{\bibfnamefont{D.~L.} \bibnamefont{Miller}},
  \bibinfo{author}{\bibfnamefont{J.~D.} \bibnamefont{Dow}},
  \bibinfo{author}{\bibfnamefont{R.~G.} \bibnamefont{Houlgate}},
  \bibinfo{author}{\bibfnamefont{G.~V.} \bibnamefont{Marr}}, \bibnamefont{and}
  \bibinfo{author}{\bibfnamefont{J.~B.} \bibnamefont{West}},
  \emph{\bibinfo{title}{The photoionisation of krypton atoms: a comparison of
  pseudopotential calculations with experimental data for the 4p asymmetry
  parameter and cross section as a function of the energy of the ejected
  photoelectrons}}, \bibinfo{journal}{J. Phys. B}
  \textbf{\bibinfo{volume}{10}}(\bibinfo{number}{16}), \bibinfo{pages}{3205}
  (\bibinfo{year}{1977}).

\bibitem[{\citenamefont{West}(1980)}]{West:80}
\bibinfo{author}{\bibfnamefont{J.~B.} \bibnamefont{West}},
  \emph{\bibinfo{title}{Progress in photoionization spectroscopy of atoms and
  molecules: an experimental viewpoint}}, \bibinfo{journal}{Appl. Opt.}
  \textbf{\bibinfo{volume}{19}}(\bibinfo{number}{23}), \bibinfo{pages}{4063}
  (\bibinfo{year}{1980}).

\bibitem[{\citenamefont{Mandal et~al.}(2017)\citenamefont{Mandal, Deshmukh,
  Kheifets, Dolmatov, and Manson}}]{PhysRevA.96.053407}
\bibinfo{author}{\bibfnamefont{A.}~\bibnamefont{Mandal}},
  \bibinfo{author}{\bibfnamefont{P.~C.} \bibnamefont{Deshmukh}},
  \bibinfo{author}{\bibfnamefont{A.~S.} \bibnamefont{Kheifets}},
  \bibinfo{author}{\bibfnamefont{V.~K.} \bibnamefont{Dolmatov}},
  \bibnamefont{and} \bibinfo{author}{\bibfnamefont{S.~T.}
  \bibnamefont{Manson}}, \emph{\bibinfo{title}{Angle-resolved wigner time delay
  in atomic photoionization: The $4d$ subshell of free and confined {Xe}}},
  \bibinfo{journal}{Phys. Rev. A} \textbf{\bibinfo{volume}{96}},
  \bibinfo{pages}{053407} (\bibinfo{year}{2017}).

\bibitem[{\citenamefont{Amusia}(1990)}]{A90}
\bibinfo{author}{\bibfnamefont{M.~Y.} \bibnamefont{Amusia}},
  \emph{\bibinfo{title}{Atomic photoeffect}} (\bibinfo{publisher}{Plenum
  Press}, \bibinfo{address}{New York}, \bibinfo{year}{1990}).


\end{thebibliography}

\end{document}